# Non-volatile bipolar resistive switching in sol-gel derived BiFeO$_3$ thin films

Chandni Kumari[1], Ishan Varun[2], Shree Prakash Tiwari[2], and Ambesh Dixit[1,3,*]

[1]*Department of Physics, Indian Institute of Technology Jodhpur, Rajasthan 342037, India*
[2]*Department of Electrical Engineering, Indian Institute of Technology Jodhpur, Rajasthan 342037, India*
[3]*Center for Solar Energy, Indian Institute of Technology Jodhpur, Rajasthan 342037, India*
[*]*ambesh@iitj.ac.in*

**Abstract:**

BiFeO$_3$ thin films are deposited on FTO coated glass substrates using a simple sol-gel deposition technique, limiting thickness about 70 nm and Ag/BiFeO$_3$/FTO RRAM devices are prepared. The devices showed low-voltage bipolar switching with the maximum $I_{on}/I_{off}$ ~ 450, and low set and reset voltages ~ 1.1 V and -1.5 V, respectively. The devices are stable against on-off cycles with ~ $10^4$ s retention time without any significant degradation. The variations in the set and reset voltages are 0.4 V and 0.6 V, respectively. We found that ohmic and trap-controlled space charge limited conductions are responsible for low and high resistance states, respectively. The resistive switching mechanism is attributed to the formation and rupturing of the metal filament during the oxidation and reduction of Ag ions for the set and reset states.

**Introduction:**

The resistive random access memory (RRAM) based non-volatile memory devices are attracting attention due to their simple device geometry, high switching speed, low power consumption, longer retention time and switching endurance in conjunction with the possibilities for scaling up towards practical applications.[1-3] This works on the principle of resistive switching, which can be realized in simple metal-insulator-metal (MIM) capacitor-like structures. The RRAM device switches between conducting and non-conducting states by manipulating external stimuli, such as electric field.[4-6] These devices are explored in numerous transition metal oxides and perovskites like $HfO_2$[7], $TiO_2$[8], $Al_2O_3$[9], $Ta_2O_5$[10], $Pr_{0.7}Ca_{0.3}MnO_3$[11], $BiFeO_3$[12], and $SmGdO_3$[13]. The onset of resistive switching behavior is attributed to the formation and rupture of conducting filaments[14-16], Schottky barriers [17], polarization induced tunneling effect [18-19] and trap charging/discharging process.[20] The formation of the conductive filament (CF) is proposed as one of the main sources for switching an RRAM device between the conducting low resistance state (LRS) and non-conducting high resistance state (HRS). The formation and rupture of the filament in RRAM is attributed to either the oxygen vacancy generation or the diffusion of active metal ions from the top or bottom electrodes into the active layer under the effect of the extremely high electric field. [21-28]

The formation of the filament is attributed to the presence of foreign atoms, especially coming from electrodes or intrinsic defects in the active material. The oxidizing metals such as Ag or Cu are used as the active top electrodes (TE) and inert metals such as Pt or W bottom electrodes (BE). The active electrode atoms may act as the source for forming a conducting filament, as foreign atoms in the active layer. The positive bias on TE leads to the oxidation of active material and the metal atoms start depositing on the counter electrode. The deposited metal atoms propagate from BE to TE because of the high electric field, and thus, form a conducting path. This causes a short circuit in the memory cell, bringing the device in ON state. Further, the filament can be dissolved by applying a negative bias on TE, causing the memory cell to return to OFF state. The device working on this model is known as conducting-bridge random access memory (CBRAM) or electrochemical metallization cells (ECM). In contrast to the foreign atoms based filament approach, high work function electrodes e.g Pt or W are used as TE for intrinsic defects mediated conducting filament. Here, the application of positive field forces the migration of oxygen vacancies in the active layer, leading to the formation of conductive filament. This will bring the memory cell in ON state, which can be reverted back to OFF state by applying the opposite polarity causing the rupture of the filament. [29] RRAM working on this model is known as valance change memory (VCM). When the device is turned ON from OFF state, then it is in Set state and when it is turned OFF from ON state, is in Reset state. These ON-OFF resistance states (RS) are reproducible, where ON and OFF states are considered as low resistance state (LRS) and high resistance state (HRS), respectively. These LRS/HRS memory devices are divided into two sub-categories: the first one is unipolar RS, where the set and reset states are determined by the amplitude of the bias voltage and is independent of the polarity and the second one is bipolar RS where set and reset states depend on bias-polarity.

Various perovskite materials are explored for RRAM application and recently bismuth ferrite is identified as a potential room temperature magnetodielectric material[30-31]. The magnetodielectric properties of bismuth ferrite may provide an additional degree of freedom to control the magnetic order parameter by external electric field or vice-versa. These thin films are deposited using pulsed laser deposition (PLD) [12,32], RF/DC sputtering [33-35] and sol-gel methods [36-37]. However, the large leakage current limits its ferroelectric properties because of the presence of unintentional oxygen vacancies. In contrast, these oxygen vacancies can be of great importance, realizing the resistive switching characteristic in

BiFeO$_3$ thin films. The formation of these vacancies depends on thin film growth process. The recent work on BiFeO$_3$ reports both unipolar and bipolar RRAM characteristics against an external applied electric field, where the ferroelectric properties and relatively large leakage current are attributed to the observed memory characteristics [32-37]. These RRAM memory devices are realized on BiFeO$_3$ thin films deposited using PLD and sputtering. However, the existence of both unipolar and bipolar resistive switching behavior in BiFeO$_3$ is not very clear. This motivated us to investigate the switching behavior in the simple and low-cost solution processed BiFeO$_3$ thin films in Ag/BiFeO$_3$/FTO RRAM device configuration. We have used silver as top contact because of its better conducting nature. FTO is used as bottom electrode because the small (~0.3 eV) work function difference between BFO and FTO, which may allow easy flow of electrons at the BFO/FTO interface, even after applying a small potential difference. The investigated device structure is the first report to the authors' knowledge, where silver (Ag) as the top and fluorine doped tin oxide (FTO) as the bottom electrodes are used to investigate the switching characteristics in BiFeO$_3$ thin films.

In this work, we report the resistive switching behavior of sol-gel derived BiFeO$_3$ thin films based Ag/BFO/FTO RRAM devices. This particular MIM structure showed relatively lower operating voltage and power consumptions. The experimental results substantiate that the switching behavior is bipolar and very stable against ON/OFF cycling. The probable mechanism for this observed RRAM characteristics is attributed to the formation and rupture of metal filaments due to the dissolution of TE under a bias voltage. These Ag/BiFeO$_3$/FTO RRAM devices showed enhanced on/off current ratio and retention time.

**Experimental details:**

The stoichiometric ratio of bismuth nitrate pentahydrate (Bi(NO$_3$)$_3$.5H$_2$O, 98% Alfa Aser) and iron nitrate nonahydrate (Fe(NO$_3$)$_3$.9H$_2$O, 98% Alfa Aser) as precursors are used in 10 ml of 2-methoxyethanol to make a 0.3 M solution. The resulted solution is continuously stirred at 80 °C till a transparent dark brown color gel is formed. This gel is aged for 24 hours at room temperature to achieve the necessary viscosity for synthesizing thin films. The gel is spin coated at 3000 revolutions per minute (RPM) for 30 seconds, followed by heating at 350 °C on a hot plate for 5 minutes in the air. This is repeated twice to achieve the desired thickness for RRAM studies. Finally, the spin-coated BFO thin films are annealed at 450 °C for 3 hours in air to achieve the desired crystallinity. The thickness of the films is measured using profilometer and is about 70 nm ± 6 nm. The schematic structures of fabricated RRAM devices are shown in Fig. 1(a). The silver "Ag" contacts are thermally evaporated on $BiFeO_3$ using a shadow mask. The geometrical dimensions for these contacts such as thickness and diameter are ~ 300 μm and 500 μm, respectively. The separation between these contacts is kept about 500 μm. RRAMs are biased with a positive potential at top electrode (TE) in conjunction with the grounded bottom electrode (BE), Fig. 1(a).

**Results and discussion:**

X-ray diffraction (XRD) measurements are carried out in the lock-coupled mode using Bruker D8 powder diffractometer at room temperature with a Cu $K_\alpha$ (λ=1.5406 A˚) monochromatic incident radiation source, operating at 40.0 kV and 40.0 mA. The scanned 2θ range is 20°-70°. The step size is 0.02° per second.. The measured XRD diffractogram is plotted in Fig 1(b) for synthesized $BiFeO_3$ thin films. $BiFeO_3$ thin films are polycrystalline and crystallized in the rhombohedral crystallographic phase structure. The results are in agreement with ICDD # 72-2035. A relative texturing along (012) direction is observed for $BiFeO_3$ thin films, Fig. 1(b), without any impurity phase.

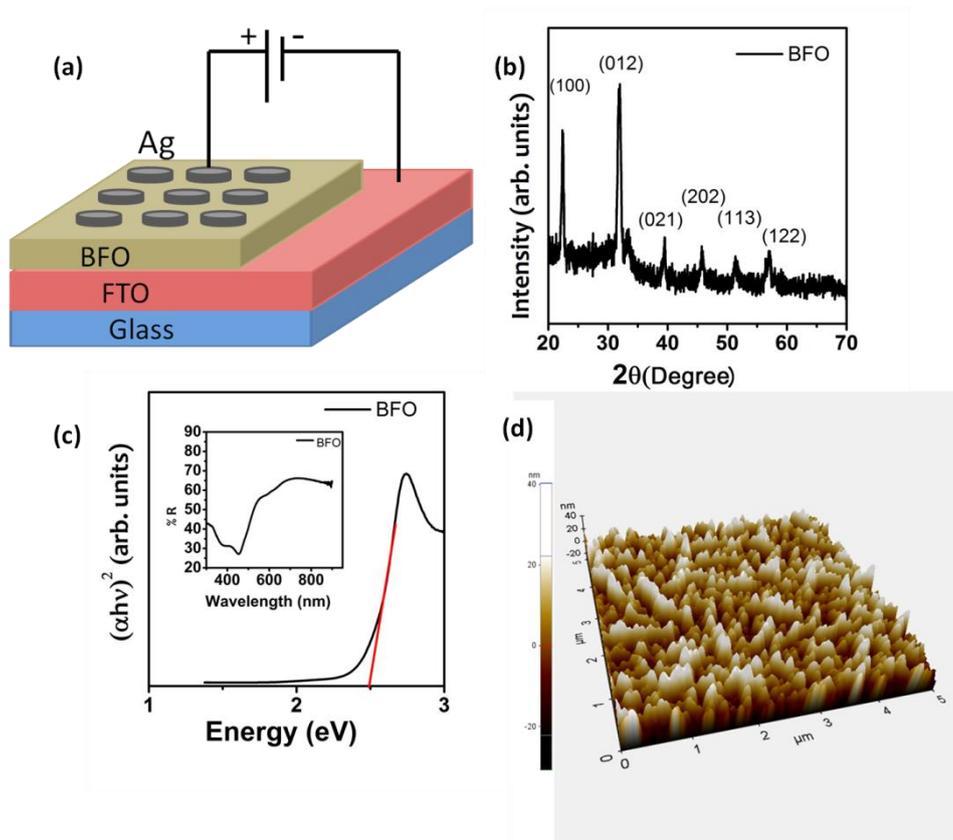

Figure1: (a) Schematic representation of Ag/BiFeO$_3$/FTO/Glass device configuration used in present investigation, (b) X-ray diffractogram, (c) ($\alpha$.h$\nu$)$^2$ versus energy (h$\nu$) and (d) atomic force microscopic image for BiFeO$_3$ thin films

The optical reflectance measurements are carried out using UV-Vis diffuse reflectance spectroscopy (DRS) and data is plotted as an inset in Fig. 1(c). The reflectance data is used to calculate absorbance '$\alpha$' using Kubelka-Munk function and ($\alpha$.E)$^2$ is plotted against energy E in Fig. 1(c). The measured optical band gap is ~ 2.47 eV, consistent with the reported literature.[38] Atomic force microscopic (AFM) measurements are carried out to probe the surface morphology and the collected image is shown in Fig. 1(d). The microstructure suggests that films are relatively smooth and root mean square (RMS) roughness is 12±0.4 nm. The observed surface roughness is attributed to the sol-gel deposition process, used for synthesizing these BFO thin films.

The resistive switching characteristics are carried out using a Keithley 4200-SCS parameter analyzer. The device configuration is shown in Fig. 1 (a) where positive potential is applied at the TE and BE is grounded. The measured current-voltage (I–V) characteristics of the Ag/BiFeO$_3$/ FTO RRAM device is shown in Fig 2(a) under -2 V→ 0 V→2 V→ 0 V→ -2 V voltage sweep. A compliance current of 10 mA is applied to avoid the hard breakdown of devices. The direction of current conduction is from bottom electrode to the top electrode through the BiFeO$_3$ film in these devices as per the applied bias profile. A high voltage is applied across the device to turn an RRAM device on for the first time and the process is called electroforming. In the electroforming process, the current rises suddenly from low to high value, leading to the resistance drop due to the formation of the primary conductive filament (CF) between TE and BE. This process is shown as an inset in Fig. 2(a), showing an abrupt increase in current at forming voltage ($V_f$) = 2.55 V. Further, when a sweep is applied from 0 to 2 V, an abrupt current increase is seen at 1.15 V in positive bias region. This increase in current switched the HRS to the LRS and the process is called the set process. The device is maintained in LRS, while voltage sweeping is carried out from 2 V to 0 V. The bias voltage is swept from 0 to -2 V without any compliance in the negative bias region, the current decreased from high to low value around -1.5 V. This sudden drop in current is attributed to the switching from LRS to HRS and is called the reset process. The device is maintained in HRS while bias is swept from -2 V to 0. The maximum $I_{on}/I_{off}$ ratio is 450 for these RRAM devices. The set and reset process depends on the polarity of the applied voltage and not on the amplitude, substantiating that investigated devices are showing bipolar resistive switching characteristics.

The reproducibility of the switching characteristics is measured upto 100 multiple cycles and results are summarized in Fig 2(b). These measurements confirm the reproducibility and stability of devices even after multiple cycles.

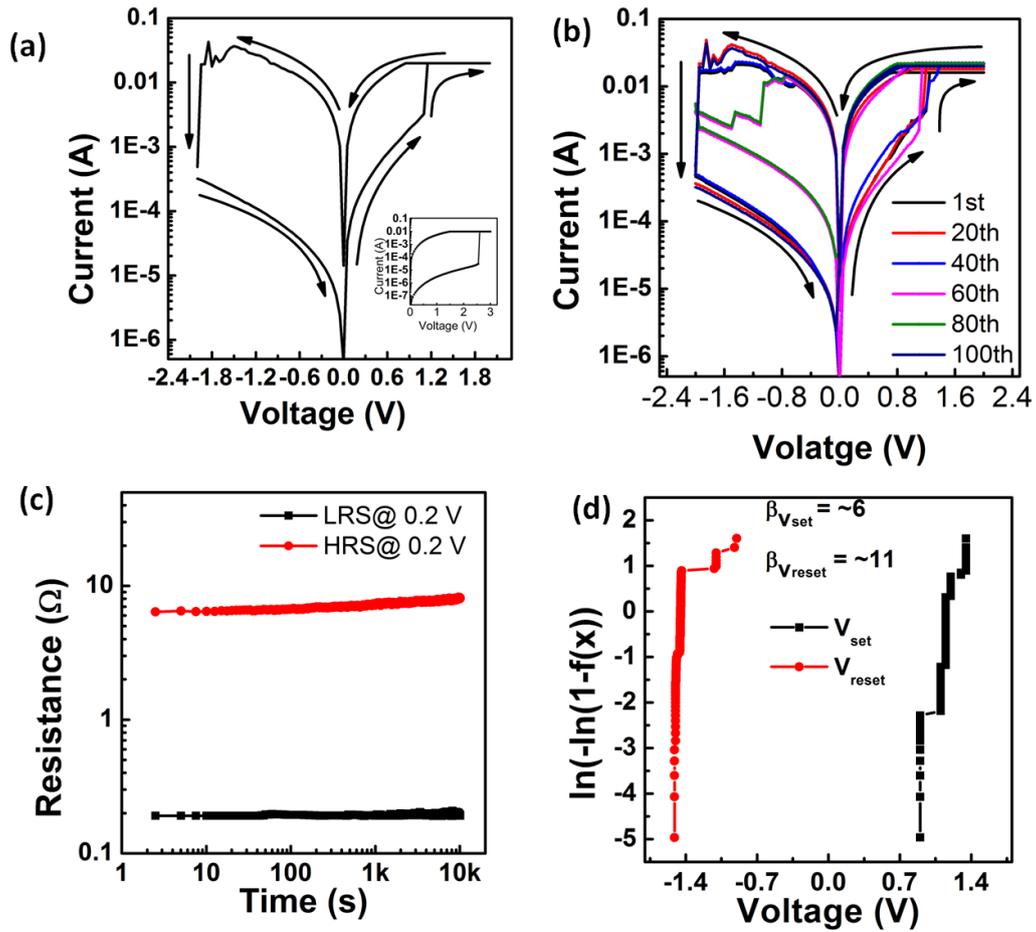

Figure 2: (a) First current-voltage (I-V) switching cycle, (b) Multiple I-V cycles up to 100 cycles with an interval of 20 cycles to ensure clarity in data, (c) retention characteristics i.e. resistance in LRS and HRS states as function of time, and (d) cumulative distribution for set and reset voltage

The retention characteristics are investigated for LRS and HRS at room temperature for $10^4$ seconds to ensure reliability and results are summarized in Fig. 2 (c). The resistance is extracted at a read voltage of 0.2 V, which showed a well-maintained storage window. The resistance values for HRS and LRSs are constant throughout the time without any significant degradation, Fig 2(c), substantiating the non-volatility of programmed logic states of the device.

Further, $V_{set}$ and $V_{reset}$ voltages are plotted in Fig 2(d) as Weibull's cumulative distribution function, defined as $\ln[-\ln\{1 - F(V)\}] = \beta \ln(V)$; where β is the shape parameter of the distribution and F(V) is the measured voltage (set or reset) function.[39] The respective values of shape parameter β are 6 and 11 for the set and reset voltage functions, respectively. The variation in reset threshold voltage $V_{reset}$ lies in the range of -0.9 V to -1.5 V and that for set voltage in 0.9 V to 1.3 V voltage window. The statistical analysis also yields the average values 1.15 V and -1.4 V for $V_{set}$ and $V_{reset}$, respectively.

The onset of current conduction mechanisms is investigated to understand the switching behaviors of $BiFeO_3$ thin films by probing the first I-V cycle in both positive and negative bias region. The logarithm of current and voltage measurements are shown in Fig 3(a) & (b). The memory cell starts with HRS initially. The conduction mechanism in RRAM memory cells, especially for HRS state, is assisted by the trap-controlled space charge limited conduction (SCLC), which consists of three regions: (i) Ohmic region (I ∝ $V^m$; m = 1), (ii) Child's law region (I ∝ $V^m$; m = 2 ), and (iii) the steep current increase region.[40-41] According to this conduction mechanism, at lower voltage (Ohmic region), the injected carriers are lower than the thermally generated charge carriers, which depends on the applied electric field and the electronic properties of the insulating materials. The injected carriers exhibit large relaxation time, which does not allow these carriers to travel through the entire film thickness. This condition leads the carriers to follow Ohm's law (m~1) and the device remains in HRS state. Further increase in the applied voltage leads to the increase in injected charge carriers, dominating over the thermally generated charge carriers. This causes filling of the available trap sites in the film. This brings Ohmic region to space charge limited region (Child's law, m~2). When all the trap sites are filled, the injected carriers are free to move along the film, leading to a sharp change in current and bringing the device in LRS state. The

Ohm's law and Child's law region are described as $J_{Ohm} = qn_0\mu\frac{V}{d}$ and $J_{Child} = \frac{9}{8}\mu\varepsilon\frac{V^2}{d^3}$ ; where q is the elementary charge, $n_0$ is the free charge carrier concentration in thermal equilibrium, µ is electron mobility, V is the applied voltage, d is the thickness of thin film and ε is the static dielectric constant[42].

The device switches from HRS to LRS state at 1.1 V during the set process and the slop of I-V curve substantiate the Ohmic conduction mechanism in this low voltage region, Fig 3(a) & (b). The different observed current conduction behavior of LRS (Ohmic conduction) and HRS (superimposed Ohmic and SCLC conduction) also support that the ON state conductivity is due to the confined filament effect rather than a homogeneously distributed charge effect. This further suggests that the active medium is much smaller than the memory cell size, providing a possibility of high-density RRAM devices.[43] The similar current conduction characteristics are observed for the negative bias region during reset process, where the transition from LRS to HRS is taking place, Fig. 3(b).

The SCLC conduction mechanism model is used to describe both type of filament formation i.e CBRAM and VCM as reported earlier[44,45]. However, resistive switching behaviour of the devices with active TE (Ag/Cu) is attributed to the formation and rupture of active electrode metal ion bridge formed between TE and BE under the application of the applied electric field[14,46,47]. These observations substantiate that observed switching response in Ag/BFO/FTO device is due to the formation and rupture of Ag metal ion bridge.

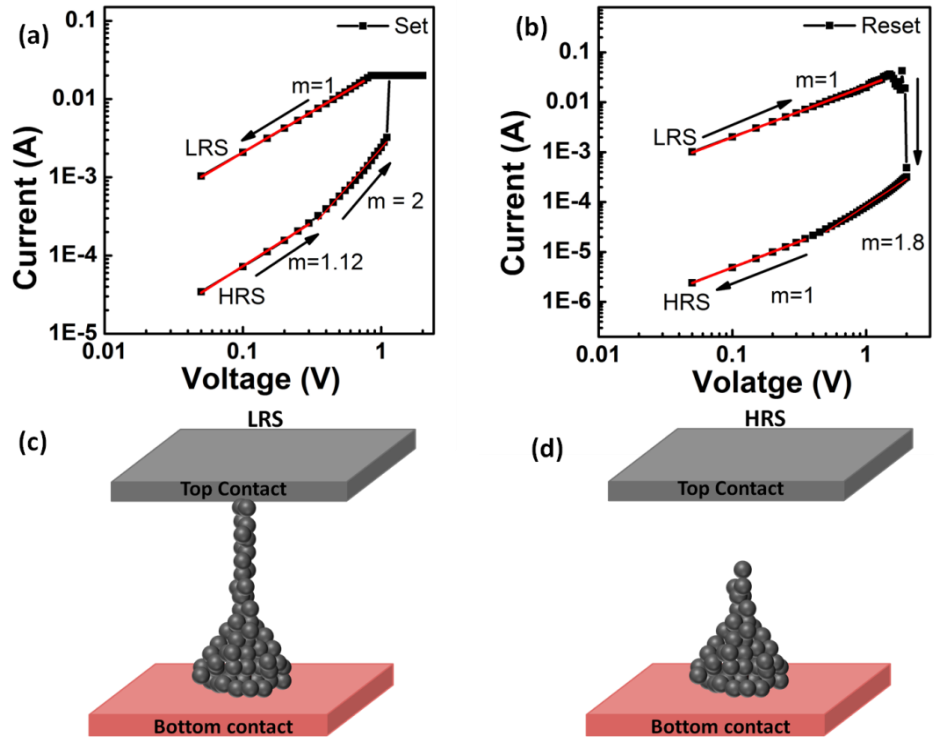

Figure 3: Current-voltage logarithmic plots for (a) set process and (b) reset process; Schematic representation of (c) Ag atom mediated conductive filament formation and (d) rupturing of the filament

Further, when a positive bias is applied on Ag TE, the metal oxidizes. As a result, $Ag^+$ cations are generated as $Ag \rightarrow Ag^+ + e^-$. These $Ag^+$ ions migrate toward FTO BE passing through BFO layer under strong electric field and these cations with electrons at BE recombine as $Ag^+ + e^- \rightarrow Ag$, forming Ag atoms. The continuous accumulation of Ag metal atoms at the cathode i.e. BE leads to the growth of an Ag conducting bridge, as shown schematically in Fig 3 (c). This conducting bridge reaches the TE at $V_{set}$ finally, causing the onset of RRAM ON state. The transition of ON state, (i.e. formation of metallic bridge) causing the Ohmic nature of LRS, as observed in Fig 3 (a) & (b). Further, on reversing the bias polarity, the dissolution of Ag atoms start along the bridge, where power dissipation is the maximum, because of Joule heating associated electrochemical reaction. This brings RRAM device back to the OFF state, as explained schematically in Fig 3 (d). [48]

## Conclusion:

In summary, BiFeO$_3$ thin films are fabricated on FTO coated glass substrates using a simple and effective low-cost solution derived process. The non-volatile bipolar resistive switching is observed for Ag/BiFeO$_3$/FTO device configuration with moderate ~ 450 I$_{on}$/I$_{off}$ ratio. The device showed large retention characteristics up to 10$^4$ s and low switching voltages ~ 1.1 V and -1.5 V for the set and reset process, respectively. The devices also showed good reproducibility up to 100 cycles. The dominant conduction mechanisms found are Ohmic and space charge controlled SCLC in LRS and HRS, respectively for both negative and positive bias regions. The LRS and HRS states are attributed to the formation and rupture of Ag conductive filaments substantiating the bipolar resistive switching performance for Ag/BiFeO$_3$/FTO RRAM devices.

## Acknowledgement:

Author Ambesh Dixit acknowledges Department of Science & Technology (DST), Government of India, through grants DST/INT/Mexico/P-02/2016 and DST/INR/ISR/P-12/2014 for carrying out experimental work.

## References:

1  R. Waser and M. Aono, "Nanoionics-based resistive switching memories.," *Nat. Mater.*, **6** [11] 833–40 (2007).

2  GI Meijer, "Who Wins the Nonvolatile," *Science (80-. ).*, **319** [5870] 1625–1626 (2008).

3  R. Waser, R. Dittmann, C. Staikov, and K. Szot, "Redox-based resistive switching memories nanoionic mechanisms, prospects, and challenges," *Adv. Mater.*, **21** [25–26] 2632–2663 (2009).

4  A. Sawa, T. Fujii, M. Kawasaki, and Y. Tokura, "Interface resistance switching at a few nanometer thick perovskite manganite active layers," *Appl. Phys. Lett.*, **88** [23] 2004–2007 (2006).

5  T. Fujii, M. Kawasaki, A. Sawa, H. Akoh, Y. Kawazoe, and Y. Tokura, "Hysteretic current-voltage characteristics and resistance switching at an epitaxial oxide Schottky junction SrRuO3/SrTi0.99Nb0.01O3," *Appl. Phys. Lett.*, **86** [1] 2003–2006 (2005).

6  X. Chen, G. Wu, H. Zhang, N. Qin, T. Wang, F. Wang, W. Shi, and D. Bao, "Nonvolatile bipolar

resistance switching effects in multiferroic BiFeO 3 thin films on LaNiO3-electrodized Si substrates," *Appl. Phys. A Mater. Sci. Process.*, **100** [4] 987–990 (2010).

7   B. Traore, P. Blaise, E. Vianello, H. Grampeix, S. Jeannot, L. Perniola, B. De Salvo, and Y. Nishi, "On the Origin of Low-Resistance State Retention Failure in HfO2-Based RRAM and Impact of Doping/Alloying," *IEEE Trans. Electron Devices*, **62** [12] 4029–4036 (2015).

8   L. Zou, W. Hu, J. Fu, N. Qin, S. Li, and D. Bao, "Uniform bipolar resistive switching properties with self-compliance effect of Pt/TiO 2 /p-Si devices," *AIP Adv.*, **4** [3] 37106 (2014).

9   H.C. D. Jana, S. Samanta, S. Maikap, "Improved Resistive Switching Memory Characteristics Using Core-Shell IrO x Nano-Dots in Al 2 O 3 / WO x Bilayer Structure," *Appl. Phys. Lett.*, **108** [11605] 1–5 (2016).

10  M.R. Park, Y. Abbas, H. Abbas, Q. Hu, T.S. Lee, Y.J. Choi, T.S. Yoon, H.H. Lee, *et al.*, "Resistive switching characteristics in hafnium oxide, tantalum oxide and bilayer devices," *Microelectron. Eng.*, **159** 190–197 (2016).

11  N. Panwar, A. Khanna, P. Kumbhare, I. Chakraborty, and U. Ganguly, "Self-Heating During submicrosecond Current Transients in Pr 0.7 Ca 0.3 MnO 3 -Based RRAM," *IEEE Trans. Electron Devices*, **64** [1] 137–144 (2017).

12  Y. Shuai, X. Ou, C. Wu, W. Zhang, S. Zhou, D. Brger, H. Reuther, S. Slesazeck, *et al.*, "Substrate effect on the resistive switching in BiFeO 3 thin films," *J. Appl. Phys.*, **111** [7] 1–11 (2012).

13  Y. Sharma, P. Misra, S.P. Pavunny, and R.S. Katiyar, "Multilevel unipolar resistive memory switching in amorphous SmGdO 3 thin film," *Appl. Phys. Lett.*, **104** [7] 1–6 (2014).

14  M. Li, F. Zhuge, X. Zhu, K. Yin, J. Wang, Y. Liu, C. He, B. Chen, *et al.*, "Nonvolatile resistive switching in metal/La-doped BiFeO3/Pt sandwiches," *Nanotechnology*, **21** [42] (2010).

15  X. Zhu, F. Zhuge, M. Li, K. Yin, Y. Liu, Z. Zuo, B. Chen, and R.W. Li, "Microstructure dependence of leakage and resistive switching behaviours in Ce-doped BiFeO3 thin films," *J. Phys. D. Appl. Phys.*, **44** [41] (2011).

16  Z.L. L. Liu, S. Zhang, Y. Luo, G. Yuan, J. Liu, J. Yin, "Coexistence of unipolar and bipolar resistive switching in BiFeO 3," *J. Appl. Phys.*, **111** 104103 (2012).

17  J.C. Li, Q. Cao, and X.Y. Hou, "Ru-Al codoping to mediate resistive switching of NiO:SnO2nanocomposite films," *Appl. Phys. Lett.*, **104** [11] (2014).

18  A.Q. Jiang, C. Wang, K.J. Jin, X.B. Liu, J.F. Scott, C.S. Hwang, T.A. Tang, H. Bin Lu, *et al.*, "A resistive memory in semiconducting BiFeO3 thin-film capacitors," *Adv. Mater.*, **23** [10] 1277–1281 (2011).

19  D. Lee, S.H. Baek, T.H. Kim, J.G. Yoon, C.M. Folkman, C.B. Eom, and T.W. Noh, "Polarity control of carrier injection at ferroelectric/metal interfaces for electrically switchable diode and photovoltaic effects," *Phys. Rev. B - Condens. Matter Mater. Phys.*, **84** [12] 1–9 (2011).

20  X. Zou, H.G. Ong, L. You, W. Chen, H. Ding, H. Funakubo, L. Chen, and J. Wang, "Charge trapping-detrapping induced resistive switching in Ba[sub 0.7]Sr[sub 0.3]TiO[sub 3]," *AIP Adv.*, **2** [3] 32165–32166 (2012).

21  C. Yoshida, K. Kinoshita, T. Yamasaki, and Y. Sugiyama, "Direct observation of oxygen movement during resistance switching in NiO/Pt film," *Appl. Phys. Lett.*, **93** [4] (2008).

22  U. Celano, Y.Y. Chen, D.J. Wouters, G. Groeseneken, M. Jurczak, U. Celano, Y. Chen, D.J. Wouters, *et al.*, "Filament observation in metal-oxide resistive switching devices Filament observation in metal-oxide resistive switching devices," **121602** [2013] 1–4 (2013).

23  C. Pearson, L. Bowen, M.W. Lee, A.L. Fisher, K.E. Linton, M.R. Bryce, and M.C. Petty, "Focused ion beam and field-emission microscopy of metallic filaments in memory devices based on thin films of an ambipolar organic compound consisting of oxadiazole, carbazole, and fluorene units," *Appl. Phys. Lett.*, **102** [21] (2013).

24  D. Lee, D.J. Seong, I. Jo, F. Xiang, R. Dong, S. Oh, and H. Hwang, "Resistance switching of copper doped MoOx films for nonvolatile memory applications," *Appl. Phys. Lett.*, **90** [12] 1–4 (2007).

25  C.B. Lee, B.S. Kang, A. Benayad, M.J. Lee, S.E. Ahn, K.H. Kim, G. Stefanovich, Y. Park, *et al.*, "Effects of metal electrodes on the resistive memory switching property of NiO thin films," *Appl. Phys. Lett.*, **93** [4] (2008).

26  K. Tsunoda, Y. Fukuzumi, J.R. Jameson, Z. Wang, P.B. Griffin, and Y. Nishi, "Bipolar resistive switching in polycrystalline TiO2 films," *Appl. Phys. Lett.*, **90** [11] 1–4 (2007).

27  C.B. Lee, B.S. Kang, M.J. Lee, S.E. Ahn, G. Stefanovich, W.X. Xianyu, K.H. Kim, J.H. Hur, *et al.*, "Electromigration effect of Ni electrodes on the resistive switching characteristics of NiO thin films," *Appl. Phys. Lett.*, **91** [8] 1–4 (2007).

28  J. Hou, S.S. Nonnenmann, W. Qin, and D.A. Bonnell, "Size dependence of resistive switching at nanoscale metal-oxide interfaces," *Adv. Funct. Mater.*, **24** [26] 4113–4118 (2014).

29  I. Valov, E. Linn, S. Tappertzhofen, S. Schmelzer, J. Van Den Hurk, F. Lentz, and R. Waser, "Nanobatteries in redox-based resistive switches require extension of memristor theory," *Nat. Commun.*, **4** 1771–1779 (2013).

30  B. Sun, Y. Liu, W. Zhao, and P. Chen, "Magnetic-field and white-light controlled resistive switching behaviors in Ag/[BiFeO$_3$/γ-Fe$_2$O$_3$]/FTO device," *RSC Adv.*, **5** [18] 13513–13518 (2015).

31  B. Sun, J. Wu, X. Jia, and P. Chen, "Photo-regulated magnetism and photoferroelectric effect in BiFeO$_3$ nanoribbons at room temperature," *Scr. Mater.*, **105** 26–29 (2015).

32  A. Tsurumaki, H. Yamada, and A. Sawa, "Impact of Bi deficiencies on ferroelectric resistive switching characteristics observed at p-type schottky-like Pt/Bi$_{1-\delta}$FeO$_3$ interfaces," *Adv. Funct. Mater.*, **22** [5] 1040–1047 (2012).

33  J. Wu, J. Wang, D. Xiao, and J. Zhu, "Resistive hysteresis in BiFeO3 thin films," *Mater. Res. Bull.*, **46** [11] 2183–2186 (2011).

34  J. Wu, X. Wang, B. Zhang, J. Zhu, and D. Xiao, "Orientation dependence of resistive hysteresis in bismuth ferrite thin films," *J. Alloys Compd.*, **569** 126–129 (2013).

35  J.J. Huang, T.C. Chang, C.C. Yu, H.C. Huang, Y.T. Chen, H.C. Tseng, J.B. Yang, S.M. Sze, *et al.*, "Enhancement of the stability of resistive switching characteristics by conduction path reconstruction," *Appl. Phys. Lett.*, **103** [4] 1–5 (2013).

36  J.M. Luo, S.P. Lin, Y. Zheng, and B. Wang, "Nonpolar resistive switching in Mn-doped BiFeO$_3$ thin films by chemical solution deposition," *Appl. Phys. Lett.*, **101** [6] 62902 (2012).

37  K. Yin, M. Li, Y. Liu, C. He, F. Zhuge, B. Chen, W. Lu, X. Pan, *et al.*, "Resistance switching in polycrystalline BiFeO3 thin films," *Appl. Phys. Lett.*, **97** [4] 1–4 (2010).

38   R. Guo, L. Fang, W. Dong, F. Zheng, and M. Shen, "Enhanced Photocatalytic Activity and Ferromagnetism in Gd Doped BiFeO3 Nanoparticles," *J. Phys. Chem. C*, **114** [49] 21390–21396 (2010).

39   J. Lawless, *Statistical Models and Methods for Life-Time Data*. 1983.

40   X. Chen, G. Wu, and D. Bao, "Resistive switching behavior of Pt/Mg0.2Zn0.8O/Pt devices for nonvolatile memory applications," *Appl. Phys. Lett.*, **93** [9] 2–5 (2008).

41   X. Chen, G. Wu, P. Jiang, W. Liu, and D. Bao, "Colossal resistance switching effect in Pt/spinel-MgZnO/Pt devices for nonvolatile memory applications," *Appl. Phys. Lett.*, **94** [3] 1–4 (2009).

42   F.C. Chiu, "A review on conduction mechanisms in dielectric films," *Adv. Mater. Sci. Eng.*, **2014** (2014).

43   Y.C. Yang, F. Pan, Q. Liu, M. Liu, and F. Zeng, "Fully room temperature fabricated nonvolatile resistive memory for ultrafast and high density memory application," *Nano Lett.*, **9** [4] 1636 (2009).

44   P.K. Sarkar, S. Bhattacharjee, A. Barman, A. Kanjilal, and A. Roy, "Multilevel programming in Cu/NiO y /NiO x /Pt unipolar resistive switching devices," *Nanotechnology*, **27** [43] 435701 (2016).

45   Y.B. Zhu, K. Zheng, X. Wu, and L.K. Ang, "Enhanced stability of filament-type resistive switching by interface engineering," *Sci. Rep.*, **7** [May] 2–8 (2017).

46   Y. Yang, P. Gao, S. Gaba, T. Chang, X. Pan, and W. Lu, "Observation of conducting filament growth in nanoscale resistive memories," *Nat. Commun.*, **3** 732–738 (2012).

47   I. Valov and R. Waser, "Comment on real-time observation on dynamic growth/dissolution of conductive filaments in oxide-electrolyte- based ReRAM," *Adv. Mater.*, **25** [2] 162–164 (2013).

48   X. Guo, C. Schindler, S. Menzel, and R. Waser, "Understanding the switching-off mechanism in Ag+ migration based resistively switching model systems," *Appl. Phys. Lett.*, **91** [13] (2007).